\documentclass{article}

\usepackage{arxiv}

\usepackage[utf8]{inputenc} 
\usepackage[T1]{fontenc}    
\usepackage{hyperref}       
\usepackage{url}            
\usepackage{booktabs}       
\usepackage{amsfonts}       
\usepackage{nicefrac}       
\usepackage{microtype}      
\usepackage{lipsum}		
\usepackage{graphicx}
\usepackage[square,numbers]{natbib}
\usepackage{doi}
\usepackage{amsmath}

\title{heterogeneous graph  based deep learning for biomedical network link prediction}

\author{ \href{https://orcid.org/0000-0002-9157-2199}{\includegraphics[scale=0.06]{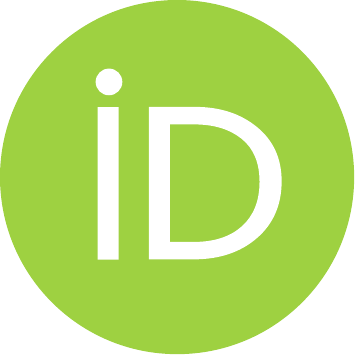}\hspace{1mm}Jinjiang Guo}, {\hspace{2mm}Jie Li}, {\hspace{2mm}Dawei Leng}, \href{https://orcid.org/0000-0001-9321-8348} {\hspace{2mm}\includegraphics[scale=0.06]{orcid.pdf} Lurong Pan*}\\
	\emph{AIDD Group}\\
	Global Health Drug Discovery Institute, Beijing, China \\
	\texttt{jinjiang.guo@ghddi.org}  and \texttt{lurong.pan@ghddi.org}  \\
}

\date{}

\renewcommand{\headeright}{}
\renewcommand{\undertitle}{}

\hypersetup{
pdftitle={GPLP},
pdfsubject={cs.AI},
pdfauthor={Jinjiang GUO, Jie Li, Dawei Leng and Lurong Pan},
pdfkeywords={Graph Neural Networks, Biomedical Network, Matrix Completion,  Link Prediction},
}

\begin{document}
\maketitle

\begin{abstract}
Multi-scale biomedical knowledge networks are expanding with emerging experimental technologies that generates multi-scale biomedical big data. Link prediction is increasingly used especially in bipartite biomedical networks to identify hidden biological interactions and relationshipts between key entities such as compounds, targets, gene and diseases. We propose a Graph Neural Networks (GNN) method, namely Graph Pair based Link Prediction model (GPLP), for predicting biomedical network links simply based on their topological interaction information. In GPLP, 1-hop subgraphs extracted from known network interaction matrix is learnt to predict missing links. To evaluate our method, three heterogeneous biomedical networks were used, i.e. Drug-Target Interaction network (DTI), Compound-Protein Interaction network (CPI) from NIH Tox21, and Compound-Virus Inhibition network (CVI). Our proposed GPLP method significantly outperforms over the state-of-the-art baselines. In addition, different network incompleteness is analysed with our devised protocol, and we also design an effective approach to improve the  model robustness towards incomplete networks. Our method demonstrates the potential applications in other biomedical networks.
\end{abstract}


\section{Introduction}
Network science is well applied in both medicine \cite{barabasi2011network} and pharmacology \cite{hopkins2008network} areas to better understand the disease interactome and system-level drug response. In the past decade, increasing multi-scale biomedicine networks \cite{campillos2008drug, morris2009autodock4, wang2014drug} have been constructed, connecting multi-level entities such like drugs, targets, activities, diseases, side-effects and omics. In  these biomedicine networks, \emph{link prediction} \cite{yue2020graph, ezzat2019computational} is one of the important network-based computing and modelling approaches for studying biomedicine relationships, which has great potency in drug repurposing \cite{zhou2020artificial}, target identification \cite{huttlin2017architecture} and biomarker discovery \cite{li2016network}. To our knowledge, researchers usually refer to recommendation algorithms \cite{luo2017network} to fulfil such link prediction tasks. There exist  two main branches of recommendation  algorithms: content-based recommendation methods \cite{pazzani2007content} and collaborative filtering models \cite{goldberg1992using}. Content-based recommendation methods take account of attacker (user) and target (item) side information as their attributes, find optimal projections between attackers and targets, and to predict future behavior of attackers or probable ratings of targets, i.e., interactions between attackers and targets. Collaborative filtering models utilize collective attacker-target links to predict missing/future interactions in the bipartite attacker-target network.

\begin{figure}
	\centering
	\includegraphics[scale=.4]{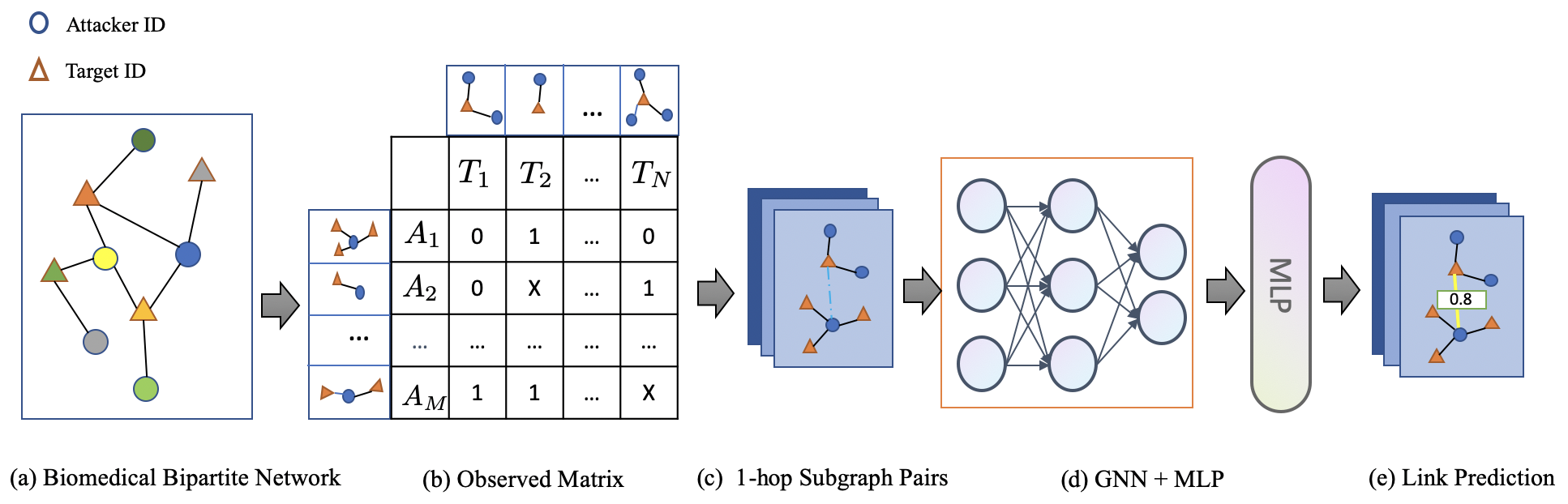}
	\caption{ Illustration of Graph Pair based Link Prediction (GPLP) framework. (a) Heterogeneous biomedical network is firstly converted into (b) observed interaction matrix  between attackers and targets. Then, extracted 1-hop subgraphs ($G^M_{a \to t}$ and  $G^N_{t \to a}$) around each attacker and target are paired as (c)  one input graph $\hat{G}_{a,t}^{M \times N}$  for training (d) GNN with MLP model. From all the training graphs $\hat{G}^{M \times N}$, GNN with MLP can learn the local graph patterns to  predict (e) link probabilities between new input attacker and target pairs. }
	\label{fig:framework}
\end{figure}

In this work, we convert collective attacker-target links from biomedical network into partially observed matrix, and  treat  the Matrix Completion (MC) task as link prediction on graphs.  Attacker-target interactions, such as Drug-Target Interaction (DTI), Compound-Virus Inhibition (CVI), etc., can be  represented as bipartite attacker-target network  in which edges denote observed links. In our proposed Graph Pair based Link Prediction (GPLP) method, we extract 1-hop subgraph around each node (attacker or target) on the bipartite network. Such 1-hop subgraphs contain \emph{local graph pattern} which is deterministic for inferring missing links, and independent on  attacker and target side information as well \cite{zhang2019inductive}.  Exhaustively around each attacker and target node pair,  we construct their 1-hop subgraphs for GNN model to extract  subgraphs latent features as representations of each attacker and target node. Such representations are used to infer the potential links via  Multi-Layer Perceptron (MLP) classifier (or regressor). Figure \ref{fig:framework} illustrates the framework of our method.

The main contributions of our work are:
\begin{itemize}
	\item We explored a novel end-to-end GNN method for predicting potential/missing links in biomedical network.
	\item Our method is independent on side information of attacker and target, simply based on their observed  connections in the network.
	\item We curated an antiviral compound-phenotype network for researchers to fight against the COVID-19 pandemic.
	\item We invented a  protocol to analyse the robustness of network based models towards real biomedical scenarios. 
\end{itemize}

This paper is organized as follows:  Section \ref{sec:relatedwork} introduces the related methods used  in complex systems. Section \ref{sec:gplpintro} gives specific description of our GPLP framework. Methods comparison results on different biomedical networks are shown in Section \ref{sec:data&exp}.    Robustness analysis of GPLP and future work  are discussed in Section \ref{sec:dis&future}.

\section{Related work}
\label{sec:relatedwork}
In this section, we briefly introduce theories of graph neural networks (GNN) and matrix completion (MC) methods used in complex systems, such as recommendation systems and biomedical networks. Also, we give a discussion on motivations of using GNN methods to fulfil MC tasks. 

\subsection{Matrix Completion}
\label{sec:mc}
Matrix Completion (MC) is a kind of task formulation \cite{candes2009exact} commonly used  in recommender systems, which converts  heterogeneous network datasets into observed matrix $\mathcal{M}$, and aims to fill the missing entries of $\mathcal{M}$. In the matrix $\mathcal{M}$,  the rows and columns  represent  users (attackers) and items (targets) respectively, each element in $\mathcal{M}$ denotes the interaction between corresponding  (attacker, target) pair coordinate. To fulfil the MC tasks, well-known Matrix Factorization (MF) methods decompose the low-rank $\mathcal{M}$ into the product of two lower dimensionality of rectangular matrices \cite{agarwal2009regression, jannach2013recommenders}. Later on, Singular Value Decomposition (SVD) is adopted in MF families \cite{kluver2014evaluating, bi2017group, pu2013understanding}, which decompose $\mathcal{M}$ into the inner product of three matrices $X$, $Y$ and $P$ (i.e., $\mathcal{M} = XPY^T$ ), where $X$ and $Y$ contain  the latent features of attackers and targets respectively, $P$ is the association matrix. The idea is to find the projections from $X$ to $Y$ by determining the optimal $P$ from $\mathcal{M}$. SVD based MF methods can integrate  latent features of attackers and targets, and therefore partially solve the \emph{cold-start} problem \cite{kluver2014evaluating}. 

Also, SVD basd MF methods successfully solved many biomedical problems, such as Drug-Target Interaction (DTI) prediction \cite{luo2017network}, Gene-Disease Association prediction.\cite{natarajan2014inductive}, etc. In the work \cite{luo2017network}, the low-dimensional  latent features of drug (attacker) and target is obtained by integrating heterogeneous data, such like drug-drug interaction, drug-side-effect, drug-disease association, drug  similarities for attacker, and protein-protein interaction, protein-disease association, protein similarities for target. The integrated features describe topological properties for each attacker and target respectively. Similarly, for predicting Gene-Disease Association \cite{natarajan2014inductive}, microarray measurements of gene expression, gene-phenotype associations of other species  and   HumanHet \cite{lee2011prioritizing} features (incorporating mRNA expression, protein-protein interactions, protein complex data, and comparative genomics) serve as latent gene (attacker) features, while disease similarities from MimMiner \cite{van2006text} and features collected from OMIM diseases \cite{hamosh2005online} are used as latent disease (target) features. As we can see, the performance of MF methods is highly dependent upon the integrated  latent features as representations  of attackers and targets. Usually, the feature construction procedure  are  manually designed, and separate from the optimization of association matrix $P$ in  MF procedure as well.

\subsection{Spatial based Graph Neural Networks}
\label{sec:gnn}
Recently, GNN based methods have demonstrated breakthroughs in many tasks regrading to network datasets \cite{wu2020graph}, such like protein structure, knowledge graph, physical systems \cite{battaglia2016interaction, sanchez2018graph}, etc. There are mainly two reasons to use GNN methods in network datasets: (1) most of the complex systems datasets are in the form of graph structure. (2) GNN methods are featured at processing topological connections among nodes, and learning graph representations \cite{wu2020graph}.

Existing GNN methods can be categorized into two types: spectral based methods and spatial based methods \cite{wu2020graph}.  For spectral based GNNs,  graphs are projected into Fourier domain where the convolution operation is conducted by computing  Laplacian eigendecomposition \cite{defferrard2016convolutional, kipf2016semi}. Due to the high computational complexity of Laplacian eigendecomposition, the Chebyshev polynomials are adopted as an approximation \cite{defferrard2016convolutional,kipf2016semi}. Spatial based GNN methods imitate convolutional neural networks (CNN) by aggregating  and updating neighborhood message from central node \cite{hamilton2017inductive, xu2018powerful, hu2019strategies, morris2019weisfeiler, ranjan2020asap, leng2021enhance}, and construct the whole graph representation through read-out function. General operations of spatial based GNN can be expressed as below: 
\begin{equation}
\label{eq:gnn}
h_v^k = \mathbf{C}^k(h_v^{k-1}, \mathbf{A}^k(\{ (h_v^{k-1}, h_u^{k-1}, e_{uv}), u \in \mathcal{N}(v) \}))
\end{equation}
where $h_v^k$ is the node feature of center node $v$  at $k$th layer, $e_{uv}$ is the edge feature between $v$ and its neighbor node $u$. $\mathcal{N}(v)$ denotes the neighborhood of node $v$, usually 1-hop neighbors.  Aggregating function $\mathbf{A(\cdot)}$ aggregates node features over neighborhood of $v$, while updating function $\mathbf{C(\cdot)}$ integrates features both from center node $v$ and its neighboring  features aggregated by  $\mathbf{A(\cdot)}$. Various mathematical operations have been adopted as $\mathbf{A(\cdot)}$ and $\mathbf{C(\cdot)}$. For instances, in work \cite{hamilton2017inductive}, mean-pooling and attention mechanisms are used as  $ \mathbf{A(\cdot)}$, whereas GRU \cite{li2015gated}, concatenation \cite{hamilton2017inductive} and summation are usually applied as $\mathbf{C(\cdot)}$.

For obtaining whole graph representation, a pooling function, namely read-out, is used at the last $K$th layer:
\begin{equation}
\label{eq:rdout}
h_G = \mathbf{R}(\{ h_v^K | v \in G\})
\end{equation}
$h_G$ means the whole representation of input graph $G$. $\mathbf{R}(\cdot)$ represents the read-out function.

Spectral based GNN methods require Fourier transform on graph,  meaning that all the  input graph samples should be static (i.e., fixed topological structure) \cite{kipf2016semi}. In contrast, spatial based GNN methods have no such restriction, and  are able to extract features on graphs with varied structures. Hence, spatial based GNNs are suitable for  the 1-hop subgraph pairs from bipartite networks in our work.

\section{Graph Pair based Link Prediction (GPLP)}
\label{sec:gplpintro}
In this part, we introduce our end-to-end Graph Pair based Link Prediction (GPLP) framework for predicting the potential links in biomedical networks. For a given bipartite biomedical network $\mathcal{G}$, we constructed the observed interaction matrix $\mathcal{M} \subset \mathbb{R}^{M \times N}$, where $M$ is the attacker number and $N$ is the target number, each row index ($a \in \mathbb{Z}^M$) and column index ($t \in \mathbb{Z}^N$)   denote the sequential identical numbers of attackers ($A^M$) and targets  ($T^N$) respectively, and each entry $m_{a,t}$ in $\mathcal{M}$ represents whole possible interaction of $(a, t)$ pair. In our case, $m_{a,t} \in [0, 1, x]$, $1$ denotes observed positive label meaning active interaction between $a$ and $t$, whereas $0$ is the observed negative label meaning the inactive interaction. Here $x$ means the unobserved (unknown) interaction waiting for prediction.

\subsection{1-hop Subgraphs Pair Construction}
\label{sec:1-hop}
For each attacker $a$ and target $t$ in the bipartite network $\mathcal{G}$, we respectively exact their 1-hop subgraphs $G^M_{a \to t}$ and  $G^N_{t \to a}$, in which the edges $E^M_{a \to t}$ and $E^N_{t \to a}$ mean the links (active interactions) of neighbouring target nodes ($\hat{t} \in G^M_{a \to t} $) around center node $a$, and neighbouring attacker nodes ($\hat{a} \in G^N_{t \to a} $) around center node $t$ respectively.  Then  for each $(a,t)$ , we pair $G^M_{a \to t}$ and $G^N_{t \to a}$ as a regrouped graph $\hat{G}_{a,t}^{M \times N}$, and label $\hat{G}_{a,t}^{M \times N}$ with $(a,t)$ corresponding $m_{a,t}$ in $\mathcal{M}$. We feed the paired subgraph dataset  to the GNN model for training or prediction. Note that after pairing the subgraphs, if $m_{a,t}$ is positive, the $(a,t)$ edge and corresponding nodes should be removed from  both $G^M_{a \to t}$ and $G^N_{t \to a}$. It is to make sure that GNN model cannot see any link for prediction in the training subgraphs pair.  In Figure \ref{fig:framework}, steps $(a)$, $(b)$ and $(c)$ show  GPLP procedures  that convert the bipartite network into GNN training dataset.  

\subsection{Graph Neural Networks Architecture}
\label{sec:gnnmodel}
The GNN backbone  we selected is the  HAG-Net \cite{leng2021enhance} developed by Global Health Drug Discovery Institute (GHDDI). HAG-Net is a variant of spatial based graph neural networks, which aggregates features by taking both  summation and maxima of neighbouring features, and thus enhances the message propagation from shallow layers to deep layers.  Specifically, for each node $v$ in graph $\hat{G}_{a,t}^{M \times N}$, its features $h_v^k$ out of HAG-Net $k$th layer is obtained  by:
\begin{equation}
\label{eq:hagnet}
h_v^k = \phi(\mathbf{concat}(h_v^{k-1}, (\sum_{u \in \mathcal{N}(v)}{h_u^{k-1}} + \max_{u \in \mathcal{N}(v)}{h_u^{k-1}} )))
\end{equation}
where $\phi(\cdot)$ is the MLP function, and $\mathbf{concat(\cdot)}$ concatenates  features of node $v$ from $(k-1)$th layer and  $k$th layer, $u$ denotes a neighbouring node in node $v$'s neighbourhood $\mathcal{N}(v)$. Then we use a mean pooling \emph{read-out} function to obtain final representation $h_{\hat{G}_{a,t}}$ of graph $\hat{G}_{a,t}^{M \times N}$. As  mentioned in \ref{sec:1-hop}, $\hat{G}_{a,t}^{M \times N}$ consists of  $G^M_{a \to t}$ and $G^N_{t \to a}$ pairs. Hence, it implies that $h_{\hat{G}_{a,t}}$  contains the final representations of both $G^M_{a \to t}$ and $G^N_{t \to a}$.

After getting the  final representation $h_{\hat{G}_{a,t}}$ of graph $\hat{G}_{a,t}^{M \times N}$,  we apply a 3-layer MLP $\Phi(\cdot)$ as functional head to predict interaction (link) probability $\hat{m}_{a,t}$:
\begin{equation}
\label{eq:mlphead}
\hat{m}_{a,t} = \Phi(h_{\hat{G}_{a,t}})
\end{equation}

\subsection{Loss Function and Optimization}
Since the link prediction in our case  is  a binary classification task, i.e. predicting active or inactive interaction, we adopt cross-entropy \cite{murphy2012machine} as our loss function, which can be described as:
\begin{equation}
\label{eq:loss}
\mathcal{L}_{a,t} = -(m_{a,t}\log{\hat{m}_{a,t}} + (1- m_{a,t})\log{(1-\hat{m}_{a,t})})
\end{equation}
where $m_{a,t}$ is the observed interaction, and $\hat{m}_{a,t}$ is the predicted link probability.
For optimization on model parameters, we tried SGD \cite{kiwiel2001convergence}, Adam  \cite{kingma2014adam} and Adabelief  \cite{zhuang2020adabelief} as optimizers, and find that Adabelief gives the best optimized  model with highest evaluation performance and most stable convergence during training and testing.

\section{Dataset and Experiments}
\label{sec:data&exp}
We conducted experiments on three heterogeneous biomedical networks: Drug-Target Interaction (DTI) dataset from DTINet \cite{luo2017network}, National Toxicology Program (NIH) Tox21 challenge dataset \cite{mayr2016deeptox}, and GHDDI constructed Compound-Virus Inhibition (CVI39) Dataset. We trained our GPLP models on each dataset respectively, and tuned each model's parameters with different protocols. In details, we took a 80-20 split on datasets of DTI and CVI39 respectively, where 80\% of dataset was used for training and the remaining 20\% was used for testing. For model training on NIH Tox21 dataset, we followed the 5-cross validation protocol from the baseline work \cite{feinberg2018potentialnet}. We took a random shuffle on each dataset for ensuring each input data sample give an independent change on the model in each training batch. Also, we resmapled the training positive and negative  data samples, balancing their number ratio to 1:1, it avoids the model's skewness towards the class with majority number during training. For the hyperparameters configuration, we followed the default settings of HAG-Net and selected Adabelief optimizer with learning rate $\eta = 0.001$, $\xi = 10^{-16}$ , $(\beta_0,  \beta_1) = (0.9, 0.999)$ and weight decay $\lambda = 0.0001$. The training batch size and epoch number were fixed to 256 and 1,000 for  all datasets. This work is publicly available  at our website  \textbf{COVID-19: GHDDI Info Sharing Portal}: \url{https://ghddi-ailab.github.io/Targeting2019-nCoV}

\subsection{DTI Dataset from DTINet}
We selected the heterogeneous network constructed in the work \cite{luo2017network}. The network includes 12,015 nodes and 1,895,445 edges in total, for predicting missing drug-target interactions. It incorporates 4 types of nodes (i.e., drugs, proteins, diseases and side-effects) and 6 types of edges (i.e., drug-protein interactions, drug-drug interactions, drug-disease associations, drug-side- effect associations, protein-disease associations and protein-protein interactions). Since our GPLP framework is independent on node's side information, only the topological information of the network is used for our method. Comparison with other counterparts, our GPLP model constantly outperforms in terms of AUROC and AUPR, reaching up to 95\% and 93\% respectively. The AUROC and AUPR we computed are the harmonic average between positive and negative performances. Our GPLP model performs nearly 5\% higher than DTINet in terms of AUROC, and achieves an approximate  AUPR. GPLP also beats other Random Walk based methods (e.g., HNM\cite{wang2014drug}), see details in Table \ref{tab:DTIcomp}.

\begin{table}
	\caption{Different methods comparison on DTI dataset. Our GPLP outperforms other state-of-the-art methods for DTI prediction.}
	\centering
	\begin{tabular}{ccccccc}
		\toprule
		Methods     & BLMNII\cite{mei2013drug}     & NetLapRLS\cite{xia2010semi}	&HNM		&CMF\cite{zheng2013collaborative} 		&DTINet 		&GPLP (ours)  \\
		\midrule
		AUROC		 & 0.67  			& 0.83			&0.86		&0.87		&0.91			&\textbf{0.96} \\
		AUPR		 & 0.74				& 0.88    		&0.88		&0.86 		&\textbf{0.93}			&0.92\\
		\bottomrule
	\end{tabular}
	\label{tab:DTIcomp}
\end{table}

\subsection{NIH Tox21 Challenge Dataset}
\begin{figure}
	\centering
	\includegraphics[scale=.45]{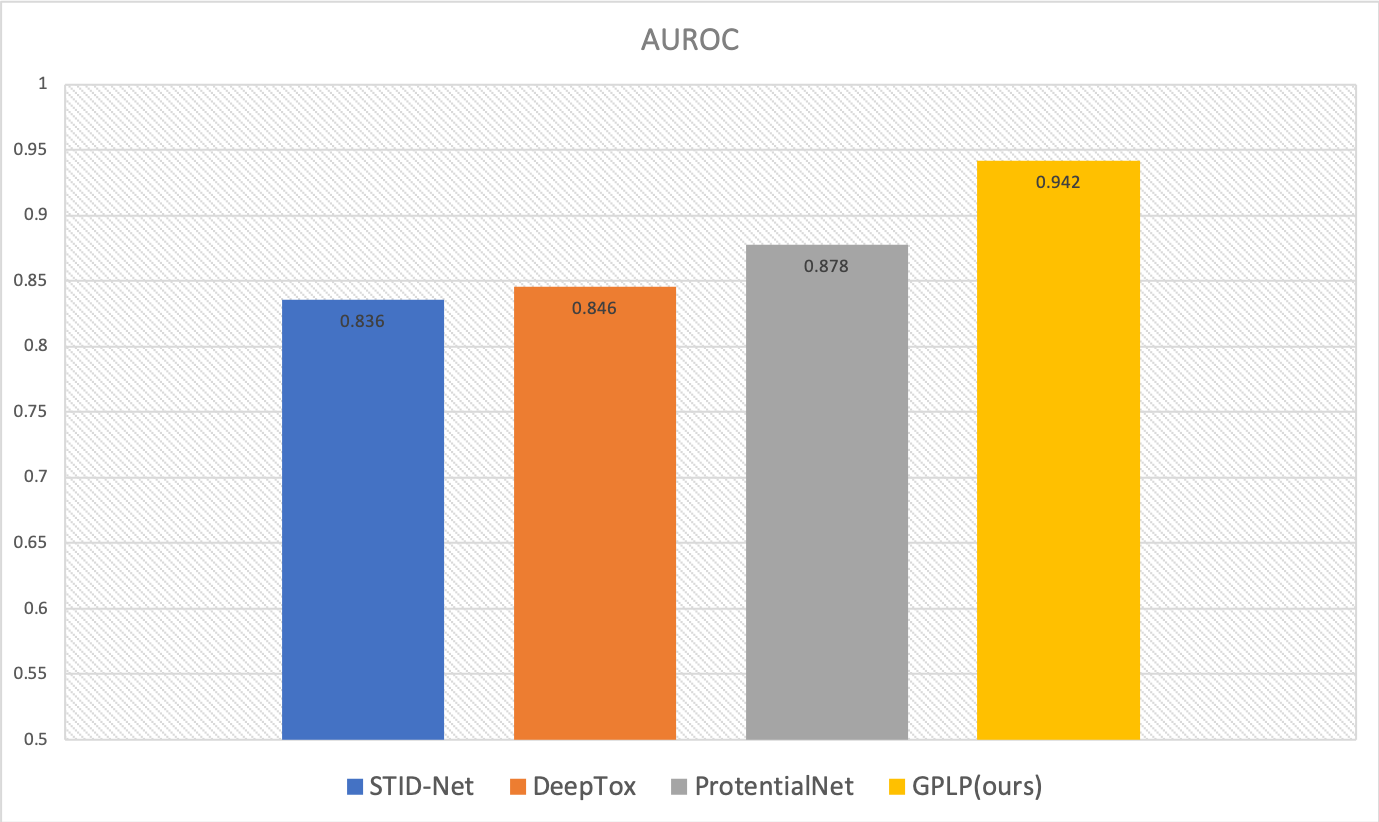}
	\caption{Different methods comparison on NIH Tox21 dataset. Our GPLP outperforms other state-of-the-art methods for toxicity prediction in terms of AUROC.}
	\label{fig:tox21comp}
\end{figure}

For NIH Tox21 Challenge Dataset \cite{mayr2016deeptox}, a dataset with 12,707 chemical compounds, which consisted of a training dataset of 11,764, a leaderboard set of 296, and a test set of 647 compounds. For the training dataset, the chemical structures and assay measurements for 12 different toxic effects were fully available at the beginning of the challenge, so were the chemical structures of the leaderboard set.To fulfil the challenge, the common methods, no matter what kind: descriptor based \cite{cao2012kernel, darnag2010support, sagardia2013new} or deep learning methods \cite{mayr2016deeptox, feinberg2018potentialnet}, focus on extracting effective representations of chemical compounds under certain toxicity task. In contrast, our prospective takes an insight  into the interactions (toxic or non-toxic) between compound (attacker) and toxicity task (target), regardless of compound chemical structures and task properties. In the experimental results, we find that our network based GPLP model significantly outperforms against the  chemical structure based deep models (e.g., DeepTox \cite{mayr2016deeptox}, ProtentialNet \cite{feinberg2018potentialnet}), achieving AUROC of 94.2\% and AUPR of 89.6\%. In addition, GHDDI also designed a chemical structure based GNN model for multi-task prediction, namely STID-Net \cite{jinjiangguo2021stidnet}, which can compete with other  chemical structure based deep methods. Hence, we added  performance  of STID-Net into comparison (see Figure \ref{fig:tox21comp}).

\subsection{CVI39 Dataset}
During COVID-19 pandemic broke-out in the year 2020, GHDDI constructed an antiviral compound-phenotype network, which collected  9,193 drugs (attacker), 39 virus families (target) and their observed  interactions of 24,135. The training dataset contains 7,737 active (IC50<=1uM) and 16,398 inactive (IC50>1uM) interactions.  For this newly constructed dataset, we conducted the experiments with our GPLP model and STID-Net as benchmarks. GPLP model achieved 95.1\% in AUROC, while STID-Net got 94.2\% in AUROC.

\section{Discussion and Future Work}
\label{sec:dis&future}
\subsection{Robustness Analysis and Improvement}
\begin{figure}
	\centering
	\includegraphics[scale=.4]{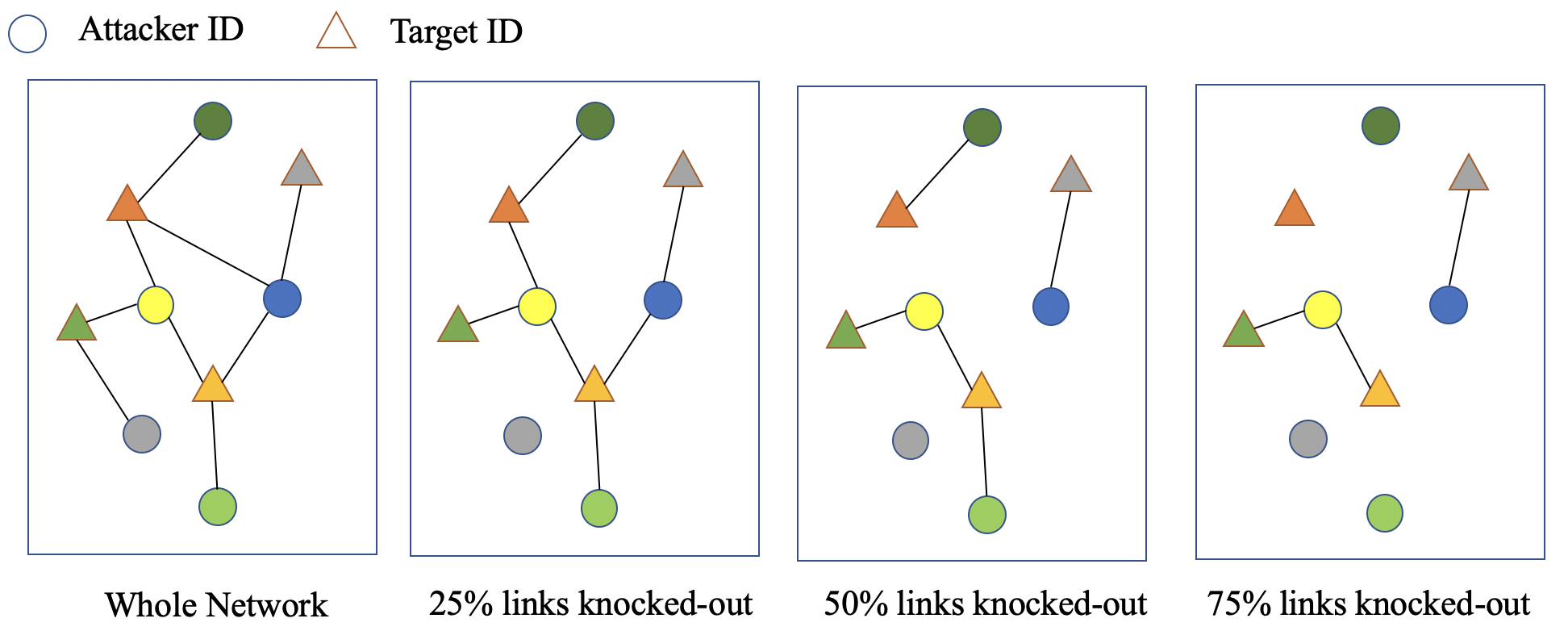}
	\caption{Different extents of knocking out links in the network to  simulate the partially observed networks in the real world scenarios.}
	\label{fig:randomknock}
\end{figure}

As in  practical biomedical problems, usually the observed links in the network are rather limited and incomplete, and thus it causes a sparse observed matrix \cite{luo2017network}. It implies that the observed links only cover a very small portion of real biomedical network  (i.e., the observed matrix is of very low-rank ). This fact results in a problem that \emph{local graph pattern} leant from training data cannot fully reveal the reality. Unfortunately, the network that covers the \emph{whole} connections  can never be obtained  in real world as our experimental dataset. To analyse our GPLP model's robustness towards real scenarios, we assume that each network we have  covers the  \emph{whole} connections, and we randomly knock-out connections in the network to simulate the partially observed network in reality (see Figure \ref{fig:randomknock}). Then, we trained GPLP model on these partial network datasets. 

\begin{figure}
	\centering
	\includegraphics[scale=.6]{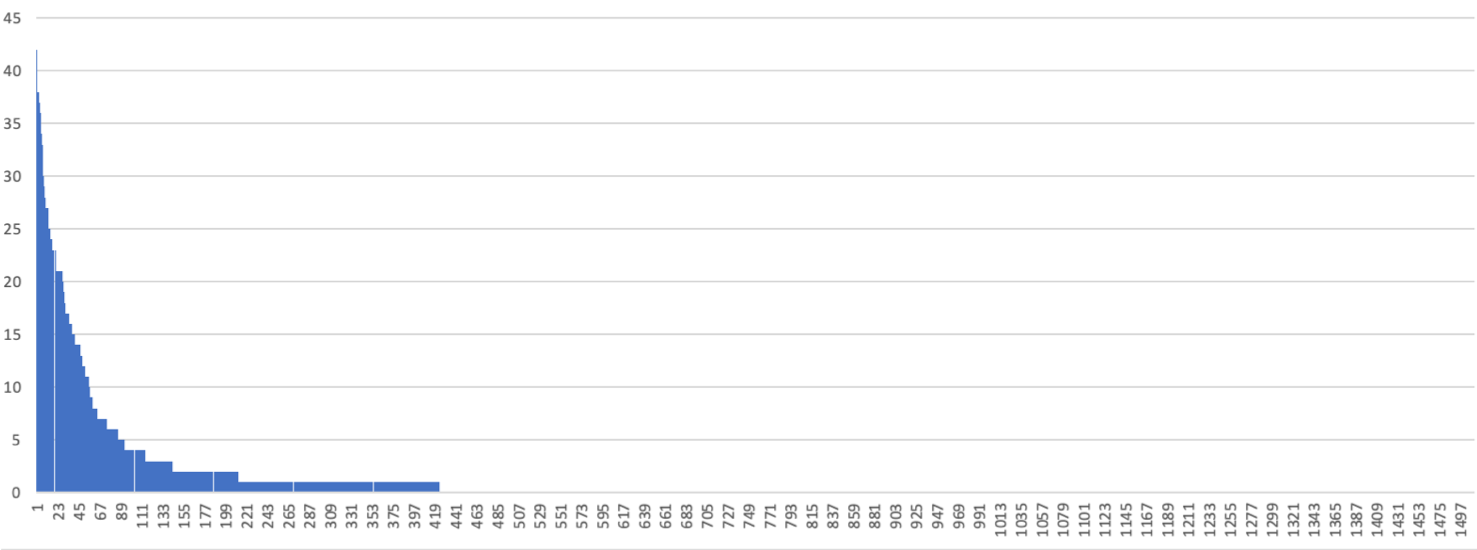}
	\caption{Attacker-target subgraph  $G^M_{a \to t}$ degree distribution on DTI dataset. A long-tail distribution.}
	\label{fig:degreedistrbdti}
\end{figure}

To implement the knocking-out protocol, we firstly convert the whole network into subgraph pairs  $G^M_{a \to t}$ and $G^N_{t \to a}$, and then randomly knock-out the edges of $G^M_{a \to t}$ and $G^N_{t \to a}$ respectively, obtaining ${G'}^M_{a \to t}$ and ${G'}^N_{t \to a}$. However, in the biomedical dataset, the degree distribution of all the $G^M_{a \to t}$ (or $G^N_{t \to a}$) usually appears as long-tail. For instance, Figure \ref{fig:degreedistrbdti} shows $G^M_{a \to t}$ degree  distribution of DTI dataset, where horizontal axis denotes the attacker (drug) ID and vertical axis means the degree of $G^M_{a \to t}$. As we can see, most of attackers have 1 or even 0 interaction with any target. In order to equally knock each subgraph, we invented  a knocking-out method by random sampling a knocking-out portion according to the mixture distributions of  $G^M_{a \to t}$ edges, which is conducted as:

\begin{equation}
\label{eq:mixtureditb}
\begin{split}
& {E'}^M_{a \to t} = \rho_\omega \oslash E^M_{a \to t}  \\
&  \rho_\omega  =  \frac{C_\Omega^\omega}{2^\Omega-1},  \;  \omega \in \Omega  
\end{split}
\end{equation}
where ${E'}^M_{a \to t}$ denotes the edges after knocking-out on original edges $E^M_{a \to t}$ of $G^M_{a \to t}$. $ \rho_\omega$ is a probabilistic function which determines the knocking-out portion of $E^M_{a \to t}$ , sign $\oslash$ means the random edge removal according to $ \rho_\omega$,  $\Omega$ is the degree of $G^M_{a \to t}$, while $C_\Omega^\omega$ means the $\omega$-combinations over $\Omega$. The knocking-out operation on $G^N_{t \to a}$ follows the same way as  $G^M_{a \to t}$ .

As was expected, the performances of GPLP models trained on knocked-out datasets declined w.r.t. AUROC and AUPR, since the topological information was constantly lost with varying degrees during training. Specifically, with this protocol, GPLP achieved AUPR of 82\% (10\% decreased ) and AUROC 96\% (approximate the same) on DTI dataset. We find that the precision of positive samples was constantly low (nearly 45\%) while their recall maintained a high level (about 80\%), it is probably because the knocking-out operates on the links that are actually positive interactions. Whereas, on NIH Tox21 dataset, the GPLP model got AUROC of 88\% ( 6\% decreased) and AUPR of 82\%( droping 8\%). The low precision (nearly 60\%) and decreasing recall (convergent to 40\%) of positive samples  appeared during training on NIH Tox21 dataset.  Even though the performance of GPLP models modestly dropped on partially knocked-out networks,  our method still gives promising results, and demonstrates the robustness in different real world scenarios.

\subsection{Conclusions and Future Work}
In this work, we have introduced our Graph Pair based Link Prediction (GPLP) framework for predicting potential/unobserved links in heterogeneous biomedical networks. GPLP model infers the linking probability between attacker and target based on the local graph patterns learnt from connectivity information of  the network. The model does not rely on the side information (chemical structures, gene sequences, etc.) of attacker or target. Our method has demonstrated an out-performance compared to state-of-the-art baselines. 

However, it is well-known that such network based methods suffer from cold-start problem \cite{kluver2014evaluating}, especially when a totally new attacker or target (e.g., SARS-CoV-2) emerges,  meaning that no  links between the new node and known nodes in the observed network. Hence, effective and efficient integration of side information into our framework become the key work in future. Nevertheless, GPLP method shows a novel perspective for revealing the underlying interaction mechanisms of complex biomedical systems. 

\bibliographystyle{unsrtnat}
\bibliography{references}  
\end{document}